# Carrier Accumulation in Organic Heterojunctions Controlled by Polarization

TAKAHASHI, Jun-ichi

8-3-5 Nagaura-ekimae Sodegaura, Chiba, 299-0246, Japan

Electronic mail: takajun@joy.ocn.ne.jp

## Abstract

The diode operation of OLEDs is characterized by the creation of a conductive space charge region in an insulating neutral dielectric region under the voltage application. It contrasts with the semiconductor diodes, where an insulating space charge region is created in a conductive charge neutral region. We proposed a heterojunction theory based on the quasi-conductor concept to give a unified view, which gave a modified Mott-Schottky equation for OLEDs (Org. Electron. **61**, 10 (2018)). Based on this theory, we proposed a new mobility evaluation method using modulus spectroscopy and reported an in-situ simultaneous evaluation of mobilities and carrier densities of carrier transporting materials in a practical OLED configuration (Jpn.J.Appl.Phys. **59**(7), 071005 (2020)). However, there were invalid handlings of the boundary conditions. We carefully reconsider the connection of the electric fields through the insulating and conductive regions based on the Dynamic Modulus Plot (DMP) analysis. It is proposed that the density of the accumulated carriers is governed by the fixed interface charges rather than the Activation of Localized Carrier Source (ALCS). Despite the errors in the previous derivation, the revised theory also gave luckily the same expression of mobility as a result, to ensure the validity of the quantitative estimation in the previous report.

## Introduction

It had been known that crystals of organic materials were able to become semiconductors since the 1940s [1]. After two innovations of the discovery of conductive polymers in 1976 [2] and the invention of ultrathin organic multi-layered devices in 1986 [3], organic electronics have achieved great success in scientific and industrial fields. As of 2020, Organic Light Emitting Diode (OLED) has achieved the greatest success as a display device in the field. Besides, Organic Field Effect Transistor (OFET) as a fundamental technology for printed electronics and Organic Solar Cell (OSC) as a low-cost flexible energy sources are being actively researched. While organic electronics are rapidly expanding in the market, our understanding of their operational dynamics is less developed. Even though our understanding of electronic states in molecular systems deepens more and more, the mysteries of device physics get deeper and deeper [4-6]. The delocalized band structure in inorganic semiconductor crystals has been reinterpreted in terms of the localized levels of the Highest Occupied Molecular Orbitals (HOMO) and the Lowest Uoccupied Molecular orbitals (LUMO) in molecules, and the device theory established by the band theory has been used to explain organic device properties. Surprisingly, despite the patchwork theory in different material groups with a different electronic nature, the performance of organic devices can be explained quantitatively to some extent [7, 8]. What are the reasons for this success? Or are we just gorging ourselves on glory under phantom prosperity? The development of organic electronics has been guided by the semiconductor theory, but the fraying of the semiconductor theory would be a burden for the next stage of developments in organic electronics.

Among the various functional elements, a diode is the most basic active circuit element that exhibits a rectifying characteristic. When a diode is made from semiconductors by sandwiching them between metal electrodes, the device acts as a variable capacitor in the reverse bias region. The capacitor stores real charges as immobile ions in the depletion layer. On the other hand, organic multi-layered devices also act as variable capacitors in the reverse bias region, which has been investigated in detail by the displacement current measurement and Impedance spectroscopy [9-17]. In a sufficiently large negative bias region, the organic layers have few intrinsic carriers and are insulators. Polarizations are induced in the organic layers and real charges are induced on both electrode surfaces. This behavior is the same as that of an ordinary capacitor made of an insulating dielectric layer sandwiched by metal electrodes. When the bias voltage is increased and carriers are injected from the electrodes, the organic layers become conductors progressively, and the capacitance changes. The device operates as a variable capacitor. However, unlike Schottky junctions, no depletion layer is formed by immobile ions in OLEDs, and the capacitance changes come from the thickening of the conductive region and the thinning of the insulating dielectric region. These differences were summarized as a quasiconductor concept [16].

We analyzed the carrier accumulation by using the Poisson equation and the continuity equation in a dielectric material and obtained the modified Mott-Schottky equation that gave a quantitative relationship between the static carrier density of the accumulation layer, the applied voltage, and the thickness of the accumulation layer [16]. Since the conductivity of the accumulation layer can be determined directly from impedance spectroscopy, the mobility can also be determined from the equation. In the reference [17], we reported an evaluation of the mobility and the carrier density from practical OLEDs. From the Dynamic Modulus Plot (DMP) analysis, it was shown that the density of the accumulated carriers was reciprocally proportional to the accumulation layer thickness and therefore the carrier density per area was independent of the accumulation layer thickness. At the same time, it was independent of the neighboring insulating layer thickness also. From these results, we assumed that the accumulated carriers were supplied from a definite external finite carrier source, which was named as Activation of Localized Carrier Source (ALCS) model, with pending the discussion about the physical picture [15].

When solving the Poisson equation and the continuity equation, we assumed the abrupt distribution approximation of the space charge density distribution similar to the Schottky contact of the semiconductors. Under the assumption of the ALCS model, the constant accumulated carrier density at the full-accumulated state can be explained by assuming that all ALCS states are ionized. However, the ALCS dynamics has no way to keep the accumulated carrier density during accumulation to be the same as that at the full-accumulated state. The quasiconductors have no intrinsic carrier density value unlike semiconductors, the carrier density during the accumulation cannot be determined a priori. The derivation in ref. (16) is no more justified. We revisited the DMP and the theory in detail and found invalid handling of the boundary conditions. In this article, we correct the errors in the theory and propose a revised assumption of the origin of the accumulation rather than ALCS. Even though another Mott-Schottky equivalent equation was obtained, the equation interrelating the accumulated carrier density and the applied voltage could be rewritten approximately as the same form as that in the previous work. The analysis in ref. (16) was found luckily to remain valid.

## Theory and discussion

We start the discussion by summarizing the concept of the quasiconductor, the difference between the semiconductor devices and the organic multi-layered electronic devices. The electronic structure of semiconductors has a band structure. The atomic orbitals not only overlap between neighboring atoms, but also have continuation among distant atoms due to the periodicity of the crystal structure, and the charges are delocalized as the Bloch state. In the case of impurity semiconductors, the dopants are ionized. At the same time, the free charges delocalize in the crystal and become carriers, giving the semiconductor a finite conductivity. The number of immobile ions and mobile carriers is balanced, making the semiconductors a charge neutral state. When electrodes are provided, a Schottky junction is formed to exclude the mobile charges in the semiconductor, leaving a region of immobile ions. The space charge potential created by the immobile ions prevents the movement of the carriers through the space charge region and switches-off the conduction [18]. On the other hand, the organic materials used in OLEDs have an amorphous structure usually. In this case, there is an overlap of molecular orbitals between neighboring molecules, but their phases cancel each other out and they are localized to individual molecules (Anderson localization) [19]. In the isolated state, there can be very few carriers in the bulk, and they are charge neutral insulators. When electrodes are provided and a voltage is applied, charges are supplied to the organic layer from the electrode, which switches-on the conduction through the organic layer. The carriers are given by the electrodes as excess charges and become space charges in the organic layer, rather than an ionization of the neutral molecules and freeing the carriers in the organic layer. The charges are spatially delocalized by repeated hopping between adjacent localized states. Note that the roles of space charges in conductors and insulators are reversed in semiconductors and organics. In semiconductors, the charge neutral state is a conductor, and the space charge state is a depletion layer and an insulator, whereas in organics, the charge neutral state is an insulator, and the space charge state is a conductor. Rectification properties can be obtained by forming heterojunctions with electrodes in both cases. However, in the case of semiconductors, rectification properties are obtained by forming an insulating depletion layer by the voltage application to the charge neutral conductor. On the other hand, in the case of organic heterojunctions, rectification properties are obtained by forming a conductive space charge layer in charge neutral insulators by the voltage application. They have a complementary relationship. We proposed a classification of quasiconductors for the organic materials in the sense that they are not conductors inherently but can be given a conductive state by appropriate manipulation, whereas semiconductors are inherent conductors that can be given an insulating state by appropriate manipulation. Quasiconductors can be defined as dielectrics with an external carrier source and finite mobility from another point of view [16]. Table 1 summarizes the differences between semiconductors and quasiconductors.

The space charge distribution in quasiconductors is described by the Poisson equation and the continuity equation so that the drift current and the diffusion current are balanced. [16].

$$\Delta \phi = -\frac{qN}{\varepsilon}. \tag{1}$$

$$0 = \sigma E - qD\nabla N. \tag{2}$$

With $q$ the elementary charge, $\varepsilon$ the permittivity, $\sigma=qN\mu$ the conductivity, $\mu$ the mobility, $N$ the carrier density, and $D$ the diffusion coefficient. For one dimensional coordinate, the following equation is obtained after integration of eq. (2).

$$A = \frac{\mu}{2D}\left(\frac{d\phi}{dx}\right)^2 + \frac{d^2\phi}{dx^2}. \tag{3}$$

This equation has general solutions as

$$\phi(x) = -\frac{2D}{\mu}\ln\left(\cosh\left(\sqrt{\frac{\mu A}{2D}}(x-x_0)\right)\right) + \phi_0(x). \tag{4}$$

$$\frac{d\phi}{dx} = -\sqrt{\frac{2DA}{\mu}}\tanh\left(\sqrt{\frac{\mu A}{2D}}(x-x_0)\right) + \frac{d\phi_0(x)}{dx}. \tag{5}$$

$$\frac{d^2\phi}{dx^2} = -\frac{qN}{\varepsilon} = A\cosh^{-2}\left(\sqrt{\frac{\mu A}{2D}}(x-x_0)\right). \tag{6}$$

Here, $\phi_0(x)$ are the solutions satisfying $\nabla\phi_0(x) = 0$. $A$ is redefined as $A = qN_0/\varepsilon$. $D$ and $\mu$ are interrelated by the Einstein relation, $1 = qD/\mu kT$. In general, this value can be larger than 1 [20,21]. The Debye length $\lambda_D$ in a dielectric material with permittivity $\varepsilon$ is defined by the following equation.

$$\zeta = \frac{qD}{\mu kT}. \tag{7}$$

$$\lambda_{D0} = \sqrt{\frac{2\varepsilon kT}{q^2 N_0}}. \tag{8}$$

$$\lambda_D = \sqrt{\frac{2\varepsilon}{q^2 N_0}\frac{D}{\mu}} = \sqrt{\zeta}\lambda_{D0}. \tag{9}$$

$\zeta$ is the measure of the deviation from the Einstein relation. In inhomogeneous dielectrics, charges are localized near the interface under the application of a potential. The extent of the localized region is expressed in terms of Debye length and becomes narrower as the charge density increases.

Next, we discuss the multi-layered devices. As noted already, the carrier behavior in organic multi-layered devices has been studied in detail by displacement current method and impedance spectroscopy [9-17]. When a bias voltage is applied to the devices, stepwise carrier injection is observed in the multi-layered device. Part of the layers is observed to shift from the insulator to the conductor at the negative bias side of the threshold voltage before the start of DC current flow. The conductive region becomes progressively thicker and saturates as the voltage increases. The accumulation voltage is the offset of the start voltage of the accumulation from the built-in potential. The voltage is known also as the giant surface potential (GSP) [11-13]. GSP represents the polarization mismatch appearing at the heterointerface. The voltage is proportional to the layer thickness, which shows that the GSP comes from a spontaneous polarization. OLEDs have a parallel-plate capacitor structure in which the stacked dielectrics are sandwiched between conductive electrodes. The organics used in OLEDs are dielectrics with variable conductivity. The carrier dynamics are represented by the charging process for the capacitors. It should be stressed that all the layers keep dielectric nature under different voltages, even though they change from insulators to conductors. It is important to pay attention to polarization and charge distribution in each layer including electrodes. Consider an OLED consisting of two layers: an electron transport layer with spontaneous polarization (layer I) and a hole transport layer (layer II). Modern practical OLEDs consist of many layers, but under proper design, the carrier transport process is simplified, and the equivalent circuit behaves similarly to HT/ET bilayer devices. Figure 1 shows the distribution of polarization and real charges under different voltages. The polarization is denoted by $P = \chi\varepsilon_0(E + E_0)$ with the spontaneous polarization $\chi\varepsilon_0 E_0$ and the electric field $E$. Under a sufficiently large negative bias, when there are no carriers in either layer, an electric field-induced polarization appears in layer I (red bars) in addition to the spontaneous polarization (yellow bars). At the same time, the same amount of the induced polarization occurs in layer II (red bars) so that the electric flux density is continuous. Corresponding to the polarization in each layer, surface charges appear on the electrodes (gray and black bars) as in Fig.1(a). When the bias voltage is increased and the electric field in layer II becomes zero, the hole exclusion due to the reverse electric field inside layer II disappears and hole accumulation begins in layer II. There remains a spontaneous polarization in Layer I (yellow bars) and surface charges on the anode surface (a gray bar) at this voltage as in Fig.1(b). When further voltage is applied, layer II becomes conductive and the layer is shortened electrically, whereas layer I remains to be insulating.

One must remember that the carriers in layer II are space charges in a dielectric that violate the charge neutrality. The potential and the charge distribution are not obvious problems. Comparing Fig.1(c) and the DMPs obtained from experiments [15-17], there are three characteristic features in the electric nature of the accumulated layer.

1. The Cole-Cole plot of modulus of the fully accumulated layer is a proper semicircle with small distortions.

2. The relaxation frequency of the accumulated layer above the full accumulation voltage does not depend on the applied voltage.

3. The relaxation frequency is reciprocally proportional to the thickness of the accumulation layer.

The first shows the following. Under a non-uniform electric field, the space charges are essentially localized in a region of about the Debye length. A typical carrier density in OLED is about $10^{16}/cm^3$ and the Debye length becomes about 30 nm, which is the same order of magnitude as the layer thickness in a typical OLED, and thus the Debye shielding effect is expected to appear. However, the semi-circle of the Cole-Cole plot in the accumulation region is a proper circle, indicating that the carrier density in the accumulation layer is highly homogeneous [15]. Therefore, the effective Debye length must be sufficiently larger than the typical thickness of OLED, indicating that the Einstein relation is broken. Then equations from eq. (4) to (6) are approximated as the following for sufficiently large $\lambda_D$.

$$\phi(x) = -\frac{qN_0}{2\varepsilon}(x^2 + mx + n). \tag{10}$$

$$E(x) = \frac{qN_0}{2\varepsilon}(2x + m). \tag{11}$$

$$N(x) = qN_0. \tag{12}$$

Here, m and n are the numerical constants that are determined from the boundary conditions. If the Debye length is sufficiently larger than the organic layer, the abrupt distribution approximation that is popular in the analysis of semiconductor junction is effective for those devices [18].

The second shows the following. We have already mentioned that OLEDs have a capacitor structure. When a voltage is applied to the capacitor, polarization occurs in the dielectric layers and real charges appear on the electrode surfaces. The first feature shows that there is no carrier localized region in the accumulation layer. The relaxation frequency is proportional to the carrier density. Therefore, the additional storage of real charges to the capacitor in this voltage range does not take place in the organic layer, even though the accumulated layer works as an electrode. The additional storage of real charges must take place on the cathode metal surface.

The third shows the following. The volumetric carrier density is reciprocally proportional to the thickness of the accumulation layer, and therefore the areal carrier density that is the integration over the thickness is constant with respect to the thickness. Here, if the origin of the accumulated carriers is due to injection from the electrode, then the accumulated carrier density will be dominated by the height of the energy barrier at the injection interface. Since the injection efficiency becomes independent of the layer thickness in such case, the thicker the layer, the more carriers are injected, and the volumetric density rather than the areal density will remain the same. The constant areal density indicates that there is some mechanism to limit the amount of carrier injection.

These three characteristics cannot be consistent with the simple capacitor model of a dielectric layer sandwiched by two metal electrodes, where real charges accumulate at the surface of the electrode metals. To account for these features, we proposed the ALCS mechanism [15], which was a phenomenological model without addressing a physical mechanism. Here, we reconsider the ALCS model and propose a new mathematical model based on the interface trap, which represents a potential balance based on the continuity of electric fluxes. It does not include the activation of the localized states explicitly, whereas the ALCS model assumes the localized states as the carrier source.

Figure 2 shows the potential and charge distribution of the model device. For simplicity, we consider a two-layer device with negative interface charges at the HT/ET interface, which accounts for the hole accumulation in the HT layer. The electron accumulation can be accounted for by the positive interface charges. The spontaneous polarization is written as only an offset of the electric field in a dielectric layer so that it is neglected here. The origins of the position and the potential are set at the HT/ET interface. We assume that there are fixed trap charges at the HT/ET interface. Due to them, there is a discontinuity of the electric field at the HT/ET interface. Holes start to accumulate in HT at the accumulation voltage *Va*, where the internal electric field in HT that prevents the flow of hole into HT disappears. At this voltage, the internal electric field in ET is compensated by the positive surface charges at the anode side as in Fig.2(a). When the bias voltage is increased, the positive surface charges at the anode decrease. At the same time, the negative charges on the opposite side decrease. When the HT/ET interface charges are fixed ones, the decrease of the negative charges is compensated by the increase of positive charges, the hole accumulation in HT in Fig.2(b). The number of accumulated carriers and the thickness of the accumulated region are linked by the applied voltage via the number of the interface charges.

Here, $\phi_{III}(t)$ is fixed to zero. When $\phi_{II}(0) \leq 0$, carriers are injected from the electrode and diffuse into the layer. The distribution of charge in a dielectric under applied voltage is governed by the competition between drift and diffusion currents. When carriers enter the layer, the space charge induces polarization in the region where the space charge has not yet spread into, and an electric field is generated in the direction that inhibits the diffusion. The electric field due to the polarization does not exceed the electric field due to the space charge. And the direction of the total electric field is always from the cathode to the HT/ET interface at the boundary between the accumulated and the not-yet-accumulated region, which is given from the potential connection condition in the discussion below. If $\phi_{II}(0)$ is non-zero (i.e., takes a finite negative value), real charges flowing from the space charge region to x=0 will appear. Since the flow of real charge continues as long as the not-yet-accumulated polarization region exists, the partially accumulated state cannot be taken. Therefore, $\phi_{II}(0) = 0$ is required for the partially accumulated region to exist. In other words, from the beginning to the end of the accumulation, it is a kind of critical state, where the inflow of diffusion current and the counterflow of drift current are balanced. The thickness of the accumulated region is the internal state quantity and the potential at both ends of the layer is the conserved quantity of the critical state. The boundary conditions are $\phi_I(-l) = V$, $\phi_I(0) = \phi_{II}(0) = 0$, $\phi_{II}(d) = \phi_{III}(d)$, $\phi_{III}(t) = 0$, and $E_{II}(d) = E_{III}(d)$. Here, $l$ is the thickness of the ET layer, $t$ and $d$ are the thicknesses of the HT layer and the accumulated region. Then, the potentials in ET (I), the accumulated region (II), and the not-yet-accumulated region (III) are written as follows.

In region I,

$$\phi_I(x) = -\frac{Vx}{l} \tag{13}$$

In region II,

$$\phi_{II}(x) = -\frac{qN_0}{2\varepsilon_1}x(x-a). \tag{14}$$

$$E_{II}(x) = \frac{qN_0}{\varepsilon_1}(2x-a). \tag{15}$$

In region III,

$$\phi_{III}(x) = -\frac{qN_0}{2\varepsilon_1}d(d-a) - E_{III}(d)(x-d). \tag{16}$$

$$E_{III}(x) = E_{II}(d) = \frac{qN_0}{2\varepsilon_1}(2d-a). \tag{17}$$

Because $\phi_{II}(0) = \phi_{III}(t) = 0$, the following relationship holds.

$$a = 2t - \frac{d^2}{t}. \tag{18}$$

There appear surface charges at the cathode to compensate for the discontinuity of the electric field at the HT/cathode interface.

$$\varepsilon_1 \phi'_{III}(t) = -\frac{qN_0}{2}(2d-a). \tag{19}$$

From the Gauss theorem, the electric field at the HT/ET interface satisfies the next equations.

$$\varepsilon_1 \phi'_{II}(0) = -qN_0 d - \varepsilon_1 \phi'_{III}(t) = \frac{qN_0(2td-d^2)}{2t}. \tag{20}$$

$$\varepsilon_1 \phi'_{II}(0) - \varepsilon_2 \phi'_I(0) = \sigma_s. \tag{21}$$

Here, $\sigma_s$ is the fixed interface charges at the HT/ET interface. The permittivity of the ET, $\varepsilon_2$ is distinguished from that of the HT, $\varepsilon_1$, here. We define the full accumulation voltage $V_f$, where the thickness of the accumulated region coincides with *t*. We obtain the accumulation equation equivalent to the Mott-Schottky equation as

$$\sigma_s \frac{(d-t)^2}{t^2} = \varepsilon_2 \frac{V-V_f}{l}. \tag{22}$$

$$qN_0 t = 2\sigma_s. \tag{23}$$

Because the thickness of the accumulated region is zero at the accumulation voltage, $V_a - V_f = \sigma_s l/\varepsilon_2$. It is shown that the areal carrier density is twice the amount of the fixed interface charges at the HT/ET interface, which is independent of the accumulation layer thickness as in Fig.2(c). It is noted that the amount of the accumulated carriers is governed by the localized states, where the localized states work as a control valve rather than a source of the carriers. It is as if the localized states activate the carries in the layer, even though the ALCS model itself is wrong. The carrier accumulation by the localized states can be considered to occupy the same position of carrier doping by the impurity doping in semiconductors as has been pointed out in ref. (15). This is another parallelism in device operation between semiconductors and quasiconductors other than table.1. The voltage difference between the start and the full state of accumulation is governed by the geometrical configuration of the device. Around the accumulation voltage, where *d<<t,* eq. (22) is written as

$$qN_0 d = \varepsilon_2 \frac{V-V_a}{l}. \tag{24}$$

From the obtained $N_0$, the mobility equation is given

$$\mu = \frac{\sigma}{qN_0} = \frac{2\pi\varepsilon_1 f}{qN_0}. \tag{25}$$

Here, $f$ is the relaxation frequency of the fully accumulated HT layer that is read from the Modulus spectrum. $\varepsilon_1$ and $\sigma$ are the permittivity and the conductivity of the layer. Equations (24) and (25) are the same equations as those derived in the ref. (16) and (17). Even though the derivation of Eq. (24) and (25) in ref. (16) was not correct, the estimation in ref. (17) remains valid by chance.

Above the full accumulation voltage, the carrier density is independent of the applied voltage. At the same time, their distribution is homogeneous. Therefore, the coefficient of the quadratic term of the potential in layer II keeps the same value as eq. (23). As $\phi_{II}(t) = 0$, the potential is written as

$$\phi_{II}(x) = -\frac{qN_0}{2\varepsilon}(x-t)(x-b), \tag{26}$$

The boundary conditions at HT/ET interface, eq. (21) remains to hold. Finally, the following equations are obtained.

$$\phi_I(x) = E(t-x). \tag{27}$$

$$E = \frac{V}{t+l}. \tag{28}$$

$$\sigma = \sigma_s - \varepsilon E. \tag{29}$$

$$b = \frac{\varepsilon E t}{\sigma_s}, \tag{30}$$

where $\sigma$ is the areal density of the charges accumulated at the cathode surface that are the stored real charges in the capacitor as in Fig.2(d). It should be stressed that the OLED works as a capacitor, all the organic layers of which are dielectrics, even though layer II is conductive. It is shown that the electric field in layer II is not shielded by the mobile carriers. The diffusion current surpasses the drift current.

The relaxation frequency of the accumulation layer keeps its value above the built-in potential until the relaxation peak of the other layer overlaps. In other words, the accumulation layer works as a regulating valve of carrier density. However, the current density in this voltage region obeys the Shockley equation with an ideality factor of 1.5 to 2 [17,22,23]. Careful theoretical investigations will be necessary to make a consistent transport model.

Finally, we point out one possibility of the nature of the accumulation. It is reported that the spontaneous polarization correlates with the molecular dipole moment in many molecules [12,13]. Within the model presented here, there is no relationship between the spontaneous polarization that governs the starting voltage of the accumulation and the interface trap charges that govern the amount of the accumulated carriers. However, the molecules during the deposition are in a very unstable state on the deposition surface and could be easily oriented under a slight forcing force. If there are surface charges on the heterointerface beforehand, the deposited molecules would feel the electrostatic field created by the interface charges as an orientation force that is proportional to their dipole moment.

Conclusion

In the previous article [16], we proposed an organic heterojunction theory based on the concept of quasiconductor. However, there were errors in the calculation. In the present article, we have corrected the derivation and revised the theory thoroughly.

The Modulus spectroscopy revealed that the carrier dynamics of OLEDs was quite different from that of the semiconductor diodes having a Schottky junction. It was shown that the roles of carriers in conductive and insulating states were switched in the organic devices and the semiconductor devices. The static behavior of carriers in organics as dielectrics has been overlooked. The polarization of each layer has a significant impact on the operation of multi-layered devices. We focus on the polarization dynamics of carrier accumulation and have shown that the carrier distribution is cleverly controlled through polarizations of the constituent layers. We derived a Mott-Schottky-equivalent equation by solving the Poisson equation and the continuity equation. Surprisingly, the equation can be approximated by the same capacitor equation derived in the previous article, even though the calculation is revised.

The concept of quasiconductor was introduced to describe this duality for semiconductors. We consider that the quasiconductor is the missing link between organic electronics and semiconductors.

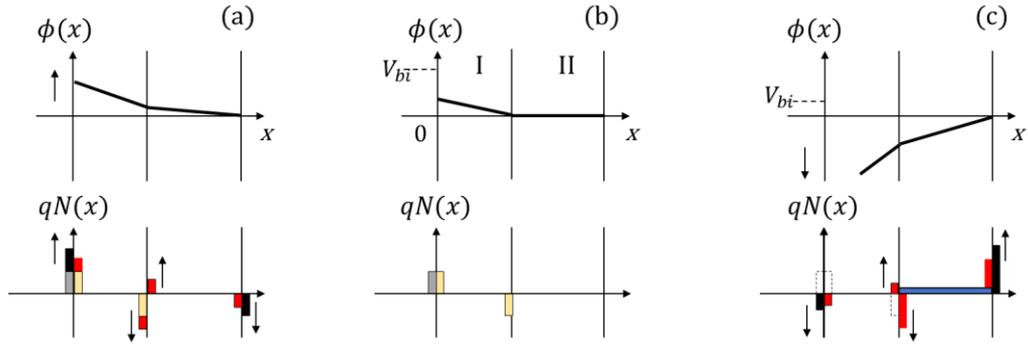

Fig.1. Potential (above) and charge distribution (below) of a two-layer device with spontaneous polarization in ET (I) and accumulation in HT (II). (a) under negative bias voltage. (b) at the accumulation voltage where the externally applied electric field balances the spontaneous polarization. (c) above the accumulation voltage. Yellow bars are the spontaneous polarization. Black and gray bars are surface charges at the electrode surface, where the black bar corresponds to the spontaneous polarization and the gray bar to the induced polarization. Red bars are the polarization induced by the applied voltage. Blue flat bar is the accumulated charges.

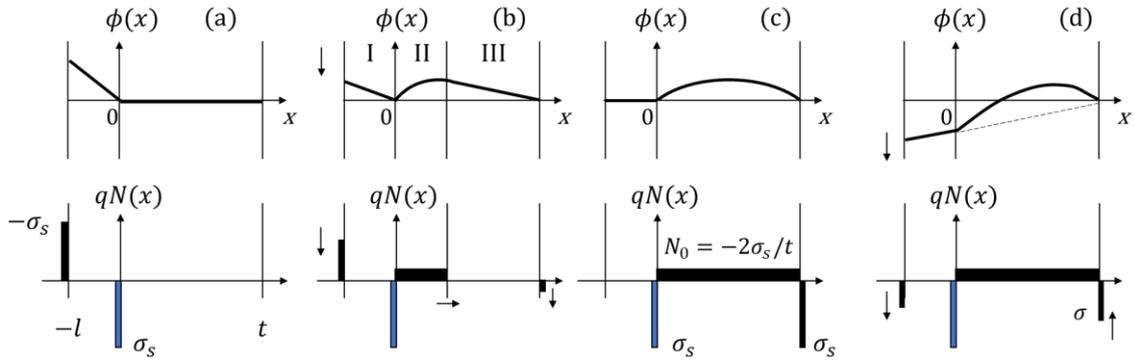

Fig.2. Potential (above) and charge distribution (below) of a two-layer device with fixed negative charges at the interface. (a) at the accumulation voltage. (b) at a finite positive bias voltage. (c) at the full accumulation voltage. (d) above the full accumulation voltage. ET layer (I), accumulated HT region (II), and not-yet-accumulated HT region (III). Black bars are real charges at the electrode surfaces and in the accumulated region, and blue bar is the fixed interface charges with a surface density of $\sigma_s$.

|  | Semiconductor | Quasiconductor |
|---|---|---|
| Charge Neutral Layer | Conductor (Delocalized Bloch states) | Insulator (Anderson Localization) |
| Space Charge Layer | Insulator (Depletion layer) | Conductor (Carrier accumulation layer) |
| Charge Delocalization | Periodic continuation of atomic orbitals | Neighboring overlap of molecular orbitals |
| Carrier Doping | Thermal excitation from mixed impurities | Electric field induced injection from outside |

Table.1 Comparison between semiconductor and quasiconductor.